\documentclass[doublecol]{epl2} 

\usepackage{amsmath}
\usepackage{cuted}
\usepackage{amsmath}
\usepackage{amssymb}
\usepackage[normalem]{ulem}

\title{Predicting Phase Ordering in Chaotic Maps and Coupled Map Lattices}
\shorttitle{Title} 

\author{Shiva Dixit\inst{1} \and Swati Chauhan\inst{2} \and
        Manish Dev Shrimali\inst{3}\thanks{E-mail: \email{shrimali@curaj.ac.in}}}
\shortauthor{ Shiva Dixit\etal}

\institute{                    
  \inst{1} Amity Institute of Integrative Sciences and Health, Amity University Haryana - Gurugram, Manesar, 122 413 Haryana, India\\
  \inst{2} Graduate School of Engineering, Nagoya Institute of Technology, Nagoya 466 8555, Japan\\
  \inst{3} Department of Physics, Central University of Rajasthan, Ajmer, 305 817 Rajasthan, India
  
}

\abstract{Coupled logistic maps exhibit collective ordering of their directional phases. As the system parameter varies, the directional phases can undergo a transition from an in-phase state to an anti-phase state, while the individual map trajectories remain chaotic. 
In this work, we propose a data-driven machine learning (ML) framework based on parameter-aware reservoir computing
(PARC) to predict order-parameter dynamics in two representative systems: a logistic
map and a two-dimensional coupled map lattice (CML). 
For the logistic map, the reservoir
is trained using only pre-crisis time series data at bifurcation parameter $\mu$ values below the attractor-merging crisis ($\mu_0 = 3.6786$).  The trained reservoir reconstructs the full bifurcation diagram and
correctly predicts the transition in the directional order parameter
$M(\mu)$,  from an ordered state
($M \approx 0$) to a disordered state ($M \neq 0$) across the crisis point. For the CML, we exploit the spatial homogeneity of
the lattice: a single reservoir is trained on the dynamics of one representative lattice
site and is then replicated
across all $L^2$ sites during prediction, where $L$=50. The replicated
reservoir correctly predicts the transition from in-phase synchronization
($\theta \approx 1$) to anti-phase clustered states ($\theta \approx 0$) at $\mu \approx 3.82$, where $\theta$
quantifies phase coherence across lattice sites. 
}

\begin{document}

\maketitle

\section{\label{sec:intro} Introduction}

Nonlinear dynamical systems often exhibit rich collective behavior, including periodicity, chaos, and abrupt transitions  \cite{strogatz2018nonlinear, ott2002chaos}. The logistic map is one of the simplest discrete-time systems that captures these phenomena and therefore serves as a model for studying nonlinear dynamics and crisis-induced changes in attractors \cite{may1976simple, ott2002chaos}. As the bifurcation parameter is varied, the logistic map undergoes a 
transition from periodic behavior, where successive iterates follow a regular pattern, to chaotic behavior. This route to chaos involves period-doubling bifurcations, followed by band merging and the eventual formation of a single chaotic band \cite{shrimali2002phase,sinha1998dynamics,sinha1997globallycoupled}.
This idea extends to coupled map lattices (CMLs), which are widely used as prototypical models for spatiotemporal chaos and collective synchronization in extended nonlinear systems \cite{kaneko1989pattern, kaneko1992overview}. In a CML, many locally coupled maps interact on a lattice, producing complex emergent behavior that includes in‑phase synchronization, anti‑phase synchronization, clustering, and spatiotemporal intermittency~\cite{chauhan2024constant}. Because of their spatially distributed and high‑dimensional nature, CMLs provide a valuable setting for investigating how local nonlinear interactions generate global ordering and phase transitions. 

Modeling and predicting such dynamics is challenging, particularly when the underlying governing equations are unavailable or when the system dimension is large \cite{abdi2024organized, zhang2023survey}. Reservoir computing (RC) provides a data-driven approach that can learn temporal evolution from time-series data with minimal training cost \cite{jaeger2001echo, lukovsevivcius2009reservoir, chauhan2025predicting, mandal2025adaptive, chauhan2023machine, yadav2025predicting, mandal2021achieving}. Recent developments in the RC field, such as the development of parameter-aware RC (PARC) architecture, further improve predictive performance by embedding the control parameter directly into the reservoir, allowing the reservoir to capture how dynamics evolve across different parameter regimes \cite{pathak2018model, vlachas2020backpropagation, sisodia2024dynamical}. Furthermore, recent studies have shown that incorporating the spatial coupling structure of the target system as an inductive bias can significantly enhance the efficiency and accuracy of RC for spatiotemporally chaotic systems \cite{chu2025incorporating}. Other works have demonstrated that a reservoir computer can be designed to emulate given coupled maps, supporting the feasibility of our approach for the coupled map lattice (CML) \cite{saha2024minimal}. Motivated by this study, we propose using a single reservoir first to learn the dynamics of a single logistic map. Later, we expand the study to utilize the spatial homogeneity of CML and utilize the single reservoir trained to learn the dynamics of a representative lattice site; the trained reservoir is then used to predict the collective behavior of the full lattice, thereby reducing computational cost while retaining predictive accuracy.

In this work, we apply the PARC approach to both a single logistic map and a $2D$ CML. For the single logistic map, we reconstruct the transition of the directional order parameter $M$ across the band‑merging crisis region, demonstrating that the reservoir captures the symmetry‑breaking signature of the transition. For the CML, we aim to predict the global phase synchronization properties of the entire high‑dimensional lattice. The trained PARC scheme successfully predicts the phase transition in terms of the in-phase synchronization order parameter $\theta$. Our results show that the proposed approach successfully predicts both the local symmetry‑breaking transition in the single map and the global synchronization behavior of the lattice, offering a computationally efficient alternative to conventional multi‑reservoir or full‑state methods.

\section{Model System}\label{model_system}

In this study, we consider a single logistic map and logistic maps coupled in a two-dimensional lattice structure.

\textbf{a. Single logistic map:} The logistic map was popularized by Robert May 
in a seminal 1976 paper as a simple discrete-time model for population dynamics 
that can exhibit surprisingly complex behavior~\cite{may1976simple}. It is defined as

\begin{equation}
\label{eq:logistic}
X_{t+1} = f(X_t), \qquad f(x) = \mu x(1-x),
\end{equation}

where $\mu \in [0,4]$ is the bifurcation (control) parameter and $t$ denotes the 
discrete time step. As $\mu$ increases beyond the Feigenbaum accumulation point 
$\mu_\infty \approx 3.5699$, the map undergoes a period-doubling route to chaos, 
eventually exhibiting a two-band chaotic attractor for $\mu \in (3.5699,\, 3.6786)$ 
before the two bands merge at the attractor merging crisis (AMC) point 
$\mu_0 = 3.6786$~\cite{Feigenbaum1978}.

To characterize the dynamics, we compute an order parameter $M(\mu)$ that captures 
the net directional tendency of successive iterates, averaged over a long time 
series of length $T$ (after discarding transients). For the single logistic map, 
$M(\mu)$ is defined as

\begin{equation}
\label{eq:M_def}
M(\mu) = \frac{1}{T} \sum_{n=1}^{T} S(n),
\end{equation}

where $S(n)$ is the directional sign of successive iterates, defined as

\begin{equation}
S(n) = \begin{cases}
+1, & \text{if } X_{n+1} - X_n > 0, \\
-1, & \text{if } X_{n+1} - X_n < 0, \\
\;\;0, & \text{if } X_{n+1} - X_n = 0.
\end{cases}
\label{eq:S_def}
\end{equation}

The case $S(n) = 0$ can occur at fixed points or within periodic windows; such 
steps are excluded from the sum in Eq.~\ref{eq:M_def} and the effective length 
of the sum is adjusted accordingly. $S(n) = +1$ and $S(n) = -1$ correspond to 
an \textit{up step} and a \textit{down step} of the iterate, respectively. The 
order parameter $M(\mu) \in [-1, 1]$; a value close to $+1$ ($-1$) indicates that 
successive iterates are predominantly increasing (decreasing), while $M(\mu) \approx 0$ 
corresponds to alternating up and down steps without a net directional bias, 
reflecting the two-band symmetry of the chaotic attractor below the AMC. 
Above the AMC, symmetry breaking causes $M(\mu)$ to deviate from zero, providing 
a sensitive indicator of the phase transition.

\textbf{b. Coupled map lattice:} The coupled map lattice (CML) was introduced 
independently by Kaneko~\cite{Kaneko1984}, Kapral, and Kuznetsov in the early 
1980s as a paradigm for spatiotemporal chaos and pattern formation in spatially 
extended systems. We consider a two-dimensional CML in which logistic maps on a 
square lattice interact via diffusive coupling of strength $\varepsilon$:

\begin{equation}
\begin{aligned}
X_{t+1}(i,j)
&= f\bigl(X_t(i,j)\bigr)
+ \frac{\varepsilon}{4}\Bigl[
f\bigl(X_t(i+1,j)\bigr) \\
&\quad + f\bigl(X_t(i-1,j)\bigr)
+ f\bigl(X_t(i,j+1)\bigr) \\
&\quad + f\bigl(X_t(i,j-1)\bigr)
- 4f\bigl(X_t(i,j)\bigr)\Bigr].
\end{aligned}
\label{eq:cml}
\end{equation}

where $f(x) = \mu x(1-x)$ is the logistic map defined in Eq.~\ref{eq:logistic}. 
The lattice has size $L \times L$ with $L = 50$, and periodic boundary conditions 
are imposed in both spatial dimensions, i.e., $X_t(L+1,j) \equiv X_t(1,j)$ and 
$X_t(i,L+1) \equiv X_t(i,1)$. The diffusive coupling scheme in Eq.~\ref{eq:cml} 
is a standard discrete Laplacian operator on the lattice. We note that while this 
coupling scheme conserves the spatial average of $f(X_t)$ exactly for a linear $f$, 
for the nonlinear logistic map it only approximately conserves the spatial mean, 
particularly in weakly coupled regimes. This has been extensively studied in the 
context of pattern formation and synchronization transitions~\cite{kaneko1989pattern, 
kaneko1992overview}.

To quantify the collective directional behavior across the lattice, we first define 
the instantaneous directional phase of each lattice site $(i,j)$ at time step $n$ 
as $S^{(i,j)}(n) = \pm 1$, defined analogously to Eq.~\ref{eq:S_def}. The 
instantaneous net directional phase of the entire lattice is then

\begin{equation}
M(n) = \frac{1}{L^2} \sum_{i=1}^{L} \sum_{j=1}^{L} S^{(i,j)}(n),
\label{eq:M_lattice}
\end{equation}

which plays the role of an instantaneous magnetization in analogy with a spin 
system, where each map is treated as a binary spin variable $S^{(i,j)}(n) = \pm 1$. 
To quantify phase synchronization (or phase-locked behavior) among the maps, 
we define the in-phase synchronization order parameter as

\begin{equation}
\theta = \frac{1}{T} \sum_{n=1}^{T} \bigl|M(n)\bigr|,
\label{eq:Theta_def}
\end{equation}

averaged over $T$ time steps after discarding transients. This quantity is 
analogous to the absolute value of the average magnetization in a spin system. 
$\theta = 1$ indicates perfect in-phase synchronization, where all maps share 
the same directional phase (all $S^{(i,j)}(n) = +1$ or all $= -1$) at every 
time step. $\theta = 0$ implies that, on average, exactly half of the maps are 
in the up phase and half in the down phase at each time step, corresponding to 
a fully disordered or anti-phase clustered state. Intermediate values of $\theta$ 
indicate partial synchronization. Previous studies have shown that for sufficiently 
large coupling $\varepsilon$, the CML exhibits synchronous directional 
phases~\cite{Pikovsky2001}, and this order parameter is particularly useful for 
detecting synchronization transitions in noisy or intermittent 
regimes~\cite{Sabe2024}.

\begin{figure*}[t]
    \centering
        \includegraphics[width=\textwidth, page=1]{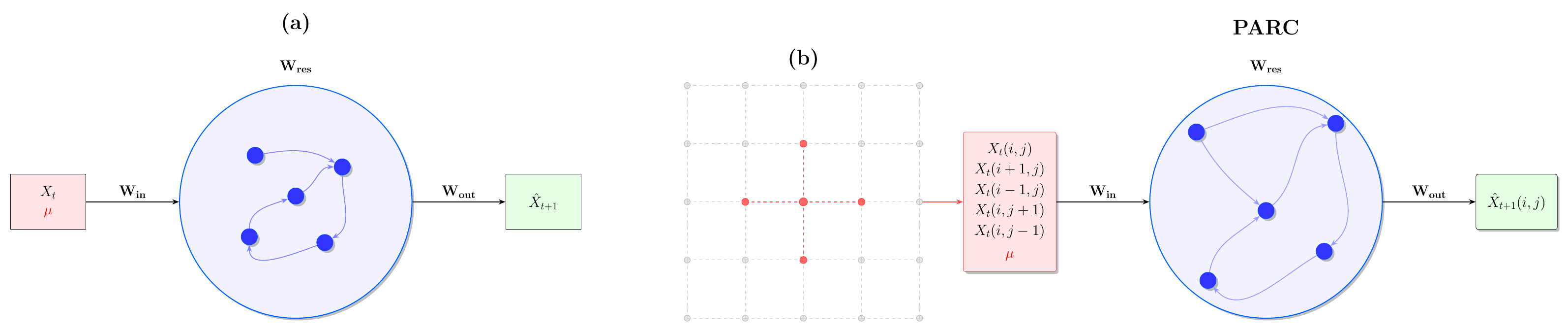} 
    \caption{Schematic diagram of the parameter-aware reservoir computing (PARC) 
    framework for (a) Single logistic map and (b) coupled map lattice (CML).}
    \label{Figure1}
\end{figure*}

\section{Reservoir Computing}\label{sec:rc}

In this work, we utilize a parameter-aware echo state network (ESN) to predict 
the dynamics of both a single logistic map and a coupled map lattice (CML). The 
primary challenge in modeling CML systems arises from their high dimensionality, 
which often necessitates a very large reservoir to learn all input variables within 
a single-reservoir framework. An alternative is a multiple-reservoir framework, 
where each reservoir captures the dynamics of a single lattice site; however, this 
approach is computationally expensive and demands substantial training resources.

To address these limitations, we propose a computationally efficient RC framework 
that requires minimal training resources. Specifically, we exploit the spatial symmetry of the lattice structure by training a single reservoir on 
the dynamics of a randomly selected lattice site and its nearest neighbors, 
analogous to the multiplexing local reservoir computing 
approach~\cite{pathak2018model, vlachas2020backpropagation}. Since all lattice 
sites obey the same local dynamical rule (Eq.~\ref{eq:cml}), a reservoir trained 
on one representative site captures the universal local dynamics of the entire 
lattice. Identical copies of the trained reservoir are then deployed during the 
prediction phase to infer the dynamics of all lattice sites independently and in 
parallel, enabling efficient forecasting across the full $L \times L$ lattice. 
This approach drastically reduces the training data required, as the single 
reservoir learns the underlying spatiotemporal dynamics from a single 
representative data stream. The schematic diagram of the proposed scheme for 
both the single logistic map and the CML is presented in Fig.~\ref{Figure1}. In this section, we describe the parameter-aware ESN framework in detail, 
covering the reservoir architecture, the training procedure, and the prediction 
procedure.

\subsection{Reservoir Architecture}

The general ESN architecture consists of three layers: an input layer, a 
reservoir (hidden) layer, and an output layer. The input layer is characterized 
by an input connection matrix $\mathbf{W}_{\text{in}}$ that feeds the input 
signals into the reservoir. The reservoir layer consists of a randomly connected 
network comprising $N_r$ nodes, where each node has its own state, and the 
collective state of all nodes is represented by the column vector $\mathbf{r}(t) 
\in \mathbb{R}^{N_r}$ at time $t$. The reservoir state is initialized to zero 
at the beginning of each training or prediction run. The state update rule is 
governed by the following leaky-integrator map:

\begin{equation}
\mathbf{r}(t+1) = (1 - \alpha)\,\mathbf{r}(t) 
+ \alpha \tanh\!\left( \mathbf{W}_{\text{res}}\,\mathbf{r}(t) 
+ \mathbf{W}_{\text{in}}\,\mathbf{u}(t) \right),
\label{eq:esn}
\end{equation}

where $\mathbf{u}(t)$ is the $m$-dimensional input signal and $\alpha \in [0,1]$ 
is the leaking rate, which controls the timescale of the reservoir dynamics. A 
value of $\alpha$ close to $1$ (as used here, $\alpha = 0.99$; see 
Table~\ref{Table 1}) implies that the reservoir state updates are nearly 
instantaneous, with minimal low-pass filtering, making the reservoir responsive 
to fast variations in the input signal characteristic of chaotic map dynamics. 
The internal connections between reservoir nodes are represented by a random, 
sparse matrix $\mathbf{W}_{\text{res}} \in \mathbb{R}^{N_r \times N_r}$, 
characterized by a connection density $k$ (percentage of nonzero entries) and 
spectral radius $\rho$ (the largest absolute eigenvalue of $\mathbf{W}_{\text{res}}$). 
The spectral radius is scaled to the desired value after random initialization 
to control the echo state property of the reservoir.

The input connection matrix $\mathbf{W}_{\text{in}} \in \mathbb{R}^{N_r \times m}$ 
couples the $m$-dimensional input vector to the reservoir. To incorporate parameter 
information into the input signal, we augment the state input with the control 
parameter $p$, so that the input at time $t$ is $\mathbf{u}(t) = [\hat{\mathbf{u}}(t),\, p]$, 
where $\hat{\mathbf{u}}(t)$ represents the $(m-1)$-dimensional state variable 
vector and $p$ is the control parameter (e.g., $p$ = $\mu$ for the logistic map). The 
matrix $\mathbf{W}_{\text{in}}$ is constructed so that the information from each 
of the $m-1$ state variables is distributed to $N_r/(m-1)$ dedicated reservoir 
nodes, while the parameter $p$ is connected to all $N_r$ reservoir nodes. Connecting the 
parameter to all reservoir nodes makes the reservoir globally aware of the 
relationship between system dynamics and its control parameter, enabling 
generalization across unseen parameter values. The input matrix $\mathbf{W}_{\text{in}}$ 
is fully dense, and its nonzero elements are drawn randomly from a uniform 
distribution $\mathcal{U}(-\sigma, \sigma)$, where $\sigma$ is the input scaling 
factor.

\subsection{Training Process}

For the single logistic map, the input signal at time $t$ is 
$\mathbf{u}(t) = [X_t,\, \mu]$, where $X_t$ is the map state and $\mu$ is the 
bifurcation parameter. The schematic of the RC scheme for the single logistic map 
is presented in Fig.~\ref{Figure1}(a). For the CML, the input signal at time 
$t$ consists of the state of a randomly selected lattice site and its four nearest 
neighbors, augmented with the bifurcation parameter:

\begin{multline}
\mathbf{u}(t) = \big[ X_t(i,j),\; X_t(i+1,j),\; X_t(i-1,j), \\
X_t(i,j+1),\; X_t(i,j-1),\; \mu \big].
\label{eq:cml_input}
\end{multline}

The schematic of the RC scheme for the CML is presented in 
Fig.~\ref{Figure1}.

During training, the reservoir is driven by the input signal across $N_p$ 
different parameter values, with $T_{\text{train}}$ time steps collected at each 
parameter value. The first $T_{\text{trans}}$ time steps at each parameter value 
are discarded as transients to allow the reservoir state to synchronize with the 
input dynamics (washout period). The numerical values of $N_p$, $T_{\text{train}}$, 
and $T_{\text{trans}}$ are listed in Table~\ref{Table 1}. After the washout, 
the reservoir states are stored and a node-wise nonlinear transformation is 
applied~\cite{gauthier2021next}:

\begin{equation}
\tilde{r}_n(t) =
\begin{cases}
r_n(t),   & \text{if } n \text{ is odd,}  \\[4pt]
r_n^2(t), & \text{if } n \text{ is even,}
\end{cases}
\label{eq:r_sym}
\end{equation}

where $n = 1,\dots,N_r$ indexes the reservoir node. This even-odd transformation 
introduces a quadratic nonlinearity into the reservoir feature space, which has 
been shown to enhance the representational capacity of the reservoir and improve 
prediction accuracy for systems with polynomial 
nonlinearities. The transformed reservoir states are 
assembled into the matrix $\mathbf{\tilde{R}} \in \mathbb{R}^{N_r \times 
N_p(T_{\text{train}} - T_{\text{trans}})}$, where each column is the transformed 
reservoir state vector at one time step. The readout layer is trained to map 
$\mathbf{\tilde{R}}$ to the target output matrix $\mathbf{Y} \in \mathbb{R}^{(m-1) 
\times N_p(T_{\text{train}} - T_{\text{trans}})}$, where each column of $\mathbf{Y}$ 
contains the true next state of the system $\hat{\mathbf{u}}(t+1)$:

\begin{equation}
\mathbf{Y} = \mathbf{W}_{\text{out}}\, \mathbf{\tilde{R}}.
\label{eq:out}
\end{equation}

The readout weight matrix $\mathbf{W}_{\text{out}} \in \mathbb{R}^{(m-1) \times N_r}$ 
is obtained analytically by minimizing the ridge regression (Tikhonov-regularized 
least squares) objective:

\begin{equation}
\mathbf{W}_{\text{out}} = \mathbf{Y}\, \mathbf{\tilde{R}}^\top 
\left( \mathbf{\tilde{R}}\, \mathbf{\tilde{R}}^\top + \beta\, \mathbf{I} 
\right)^{-1},
\label{eq:ridge}
\end{equation}

where $\beta > 0$ is the Tikhonov regularization parameter that prevents 
overfitting and ensures numerical stability of the matrix inversion. The value 
of $\beta$ is chosen by cross-validation and is listed in Table~\ref{Table 1} 
for each case.

\subsection{Prediction Process}

Once training is complete, the reservoir operates in a closed-loop (autonomous) 
prediction mode, in which the predicted output at each time step is fed back as 
the state input for the next step. To initialize the reservoir state at the 
beginning of prediction, a washout period of $T_{\text{trans}}$ steps is first 
run using a known initial condition at the new parameter value $p_{\text{new}}$, 
after which the reservoir transitions to fully autonomous prediction. The 
closed-loop update equations are:

\begin{equation}
\begin{aligned}
\mathbf{r}(t+1) &= (1 - \alpha)\,\mathbf{r}(t) 
+ \alpha \tanh\!\Bigl(\mathbf{W}_{\text{res}}\,\mathbf{r}(t) 
+ \mathbf{W}_{\text{in}}\,\mathbf{v}(t) \Bigr), \\[6pt]
\hat{\mathbf{v}}(t) &= \mathbf{W}_{\text{out}}\,\tilde{\mathbf{r}}(t), \\[6pt]
\mathbf{v}(t) &= \bigl[\hat{\mathbf{v}}(t),\; p_{\text{new}}\bigr],
\end{aligned}
\label{eq:esn_pred}
\end{equation}

where $p_{\text{new}}$ is the new parameter value for which dynamics are to be 
predicted (which may be outside the training range, enabling extrapolation across 
unseen regimes), $\hat{\mathbf{v}}(t) \in \mathbb{R}^{m-1}$ is the predicted 
state vector at time $t$, and $\tilde{\mathbf{r}}(t)$ denotes the transformed 
reservoir state as defined in Eq.~\ref{eq:r_sym}. For the CML prediction, 
identical copies of the single trained reservoir are run in parallel for each 
lattice site $(i,j)$, with each copy receiving its own local neighborhood input 
during the washout phase, after which all copies operate autonomously in closed 
loop. The complete set of hyperparameters used for both cases is listed in 
Table~\ref{Table 1}.

\section{Results}

\begin{figure*}[!ht]
    \centering
     \includegraphics[width=0.6\linewidth]{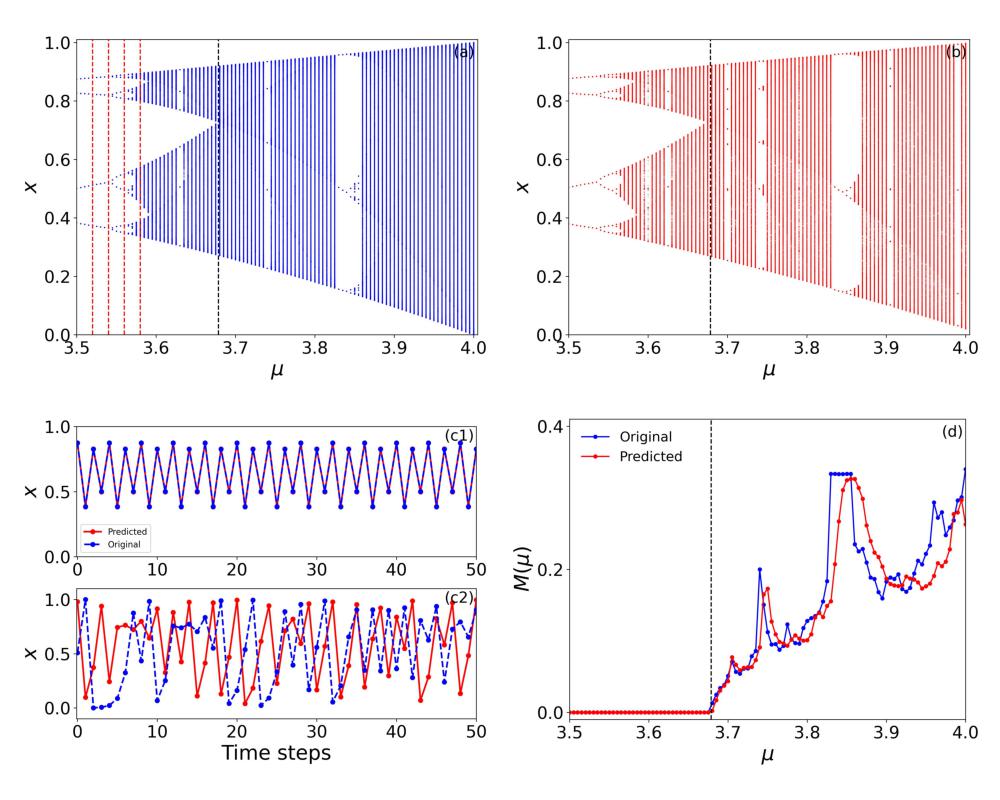}
     \caption{(a) Original bifurcation diagram of the logistic map. Vertical dashed red lines are the training parameter values. (b) Predicted bifurcation diagram of logistic map. (c1-c2) Original and predicted time series. (d) Comparison of the order parameter $M(\mu)$ for the single logistic map (average over 20 random RC realizations). The vertical dashed black line indicates the transition point at $\mu_0 = 3.6786$.}
    \label{Figure2}  
\end{figure*}

First, we apply PARC to reconstruct the bifurcation diagram of a single logistic map. To further characterize the predicted dynamics, the ``directional phase'' (which defines the sign of the difference between two consecutive iterates) is computed over a short segment of the predicted time series. The results are presented in Fig.~\ref {Figure2}. Next, we extend the PARC approach to predict the dynamics of a system of coupled logistic maps arranged in a lattice structure. A single reservoir is trained to predict the dynamics of a randomly selected logistic map within the lattice. Owing to the homogeneity of the system, replicas of the trained reservoir are used to predict the dynamics of all coupled maps across the lattice. To quantify collective behavior, we also compute an order parameter $\theta$ that characterizes phase synchronization among the maps.

\textbf{Single logistic map:}

The original bifurcation diagram of the logistic map is presented in Fig.~\ref{Figure2}(a). The single logistic map exhibits a two‑band chaotic attractor for $\mu$ approximately in the range $3.59 < \mu < 3.6786$ (just above the onset of chaos up to the merging crisis). The iterates periodically hop between these two bands, implying a sense of symmetry (two‑band symmetry). The two bands merge at $\mu_0 = 3.6786$, leading to symmetry breaking. This phenomenon is often called the attractor merging crisis (AMC).

We aim to use the PARC scheme to predict the dynamics of a single logistic map as the bifurcation parameter $\mu$ varies. The RC scheme is trained with the dynamics at $\mu \in [3.52, 3.54, 3.56, 3.58]$, selected prior to the crisis point. The vertical red lines in Fig.~\ref{Figure2}(a) represent the training values of $\mu$. The corresponding time series for training at each $\mu$ value is collected using Eq.~\ref{eq:logistic} after discarding the initial transient behavior.

The trained reservoir is employed to predict the bifurcation diagram of a single logistic map for varying $\mu$ values. The predicted bifurcation diagram is shown in Fig.~\ref{Figure2}(b). The parameter $\mu$ is varied from $3.5$ to $4.0$ with a step size $\Delta\mu = 0.01$. In addition, we plot the predicted time series $X$ as a function of time step $n$ for two distinct values of $\mu$, chosen from regions before and after the crisis point.

Furthermore, we compute the order parameter $M(\mu)$ using Eq.~\ref{eq:M_def} for the time series predicted at $\mu \in [3.5,4.0]$ with $\Delta\mu = 0.01$. We compute $M$ for 20 random realizations of the RC scheme (different random initializations of the reservoir weights), and the corresponding average value is plotted in Fig.~\ref{Figure2}. We also plot the original order parameter $M_{\text{orig}}(\mu)$. It is observed that for $\mu < \mu_0$, the net directional phase exhibits an ordered arrangement with each up phase followed by a down phase [see time series c1 in Fig.~\ref{Figure2}]. As a result, the net directional phase becomes zero, i.e., $M(\mu) = 0$, and this behavior extends up to $\mu = \mu_0$. This ordered arrangement breaks for $\mu > \mu_0$, leading to a net nonzero directional phase $M(\mu) \neq 0$. This signifies a transition from an ordered to a disordered arrangement in the single logistic map, marked by the order parameter $M$ increasing from $0$ to a nonzero positive value. The presence of flat plateaus with constant $M(\mu)$ demonstrates the existence of periodic windows (see Fig.~\ref{Figure2}). Notably, these plateaus are not prominently observed in the predicted order parameter because $M$ is averaged over 20 random realizations. A slight shift in the predicted periodic windows across different random realizations leads to smoothing of the plateaus. However, the transition point is correctly identified, and the predicted $M$ follows the original $M$ closely. The hyperparameters for this case are listed in Table~\ref{Table 1}.

\textbf{Coupled map lattice:}

\begin{figure*}[!ht]
    \centering
     \includegraphics[width=1.0\linewidth]{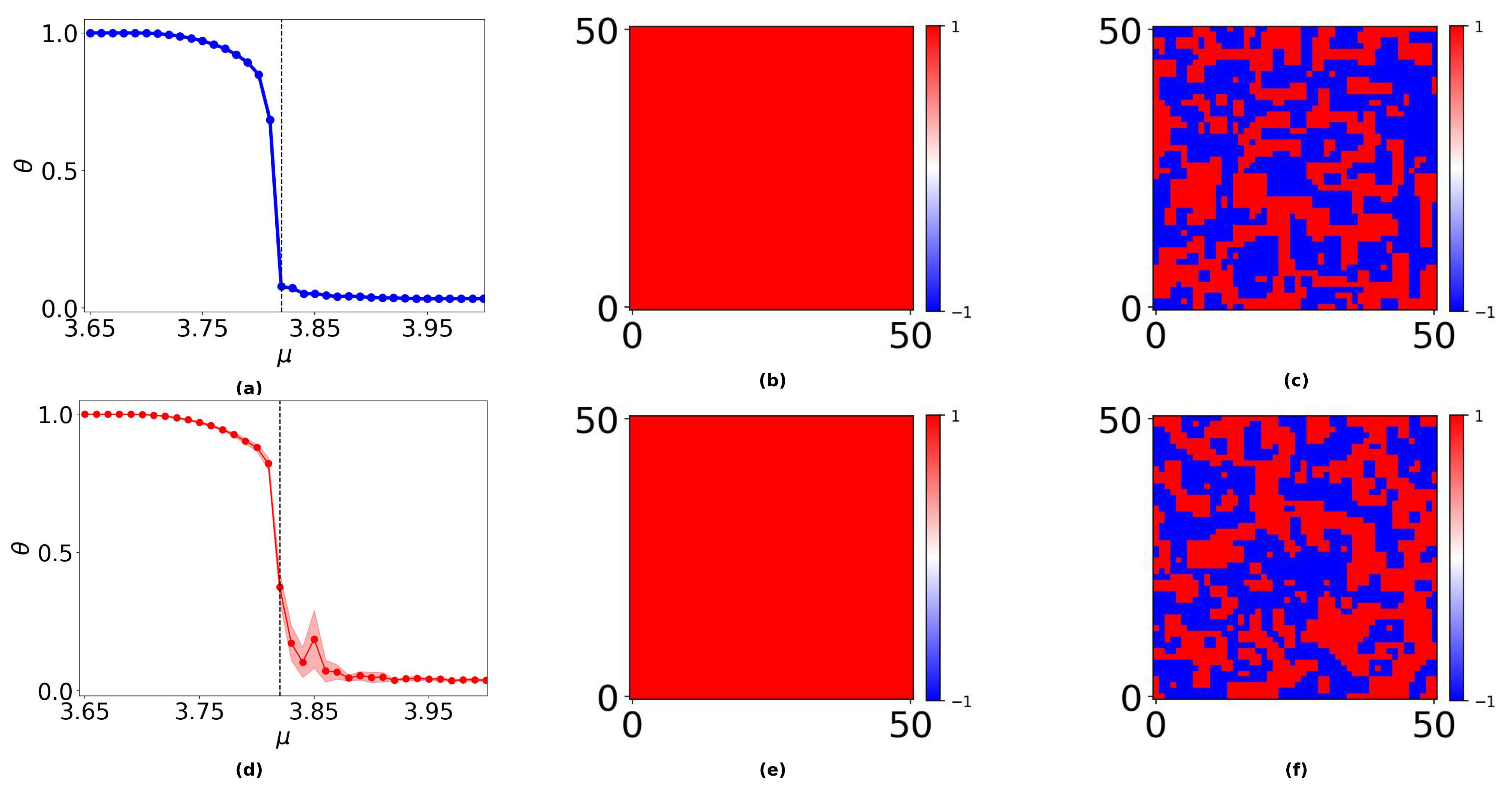}
    \caption{Comparison of the collective dynamics of the coupled map lattice (CML) obtained from the original and predicted dynamics. (a) The order parameter $\theta$ as a function of the control parameter $\mu$ for the original CML. (d) The corresponding prediction of $\theta$ obtained from the proposed data-driven model, with the shaded region representing the standard deviation over 20 independent realizations. (b,c) Spatial snapshots of the original CML for representative values of $\mu$ corresponding to the in-phase ($\mu$=3.65) and anti-phase ($\mu$=3.90) regimes, respectively. (e,f) Corresponding spatial snapshots from the predicted CML dynamics at the same parameter values, demonstrating the ability of the model to reproduce the respective in-phase and anti-phase collective states. Each lattice snapshot represents a $50\times50$ CML.}

    \label{Figure3}  
\end{figure*}

We consider a lattice of size $L = 50$, comprising $50 \times 50$ logistic maps. The coupling strength is fixed at $\varepsilon = 0.4$. The governing equations of the CML are described in Eq.~\ref{eq:cml} of Sec.~\ref{model_system}. The original order parameter $\theta$ for the CML is computed following the definitions in Sec.~\ref{model_system}. Fig.~\ref{Figure3} demonstrates that for $\mu < 3.82$, the system exhibits strong in‑phase synchronization with $\theta \approx 1$. In this regime, all maps display an up phase at time $n$ followed by a down phase at time $n+1$, resulting in synchronous oscillations of the directional phases across the lattice. Notably, as $\mu$ approaches $3.82$, a transition from strong in‑phase synchronization to anti‑phase synchronization occurs. In this anti‑phase state, the lattice contains numerous clusters that alternate between up and down phases. Consequently, the net order parameter approaches zero, $\theta \approx 0$. It is noteworthy that even in the phase‑locked state, the values of the maps $X_n(i,j)$ exhibit chaotic dynamics (see Fig.~\ref{Figure3}).

We now implement the proposed RC scheme for the CML as outlined in Sec.~\ref{model_system}. A single reservoir is trained on the data of a randomly chosen lattice site. The training is performed with time series generated at $\mu = [3.56, 3.58, 3.60, 3.62]$. The order parameter $\theta$ for the predicted time series is computed for $\mu \in [3.65, 4.0]$ with $\Delta\mu = 0.01$, following the definition in Eq.~\ref{eq:Theta_def} of Sec.~\ref{model_system}. The predicted $\theta$ obtained over 20 random realizations of the proposed scheme is presented in Fig.~\ref{Figure3} along with the original $\theta$. The vertical red lines in Fig.~\ref{Figure3} indicate the $\mu$ values used for training.

The results demonstrate that the proposed scheme captures the transition point (the $\mu$ value where the system shifts from in‑phase to anti‑phase synchronization). Notably, the reservoir is trained only on data before this transition point, implying that it has no information about the transition region. Additionally, we plot the predicted directional phases of all maps in the lattice, which shows that for $\mu$ values before the transition, all maps share the same directional phase. Beyond the transition, clusters of maps with the same directional phase emerge. Finally, Fig.~\ref{Figure3} presents the predicted values of the maps in the lattice at $\mu = 3.65$ (which is $< 3.82$), showing chaotic behavior. 

\begin{table}
\caption{ESN hyperparameter values optimized for the parameter prediction of the single map and CML}
\label{Table 1}
\begin{center}
\begin{tabular}{|p{3.4cm}|c|c|}
\hline
\textbf{Hyperparameters}  & \textbf{Logistic Map} & \textbf{CML}\\
\hline
         Number of Nodes ($N_{res})$&  400 & 600  \\
         \hline
         leakage rate ($\alpha$)& 0.99 & 0.99  \\
        \hline
        Input scaling factor ($\sigma$) & 0.2  & 0.2 \\
        \hline
         spectral radius ($\rho$)&  0.2 & 0.5  \\
        \hline
         density of $\mathbf{W}_{res}$ & 0.25 & 0.16  \\
        \hline
         regularization parameter ($\beta$)& $1.0 \times 10^{-8}$ &  $4.4 \times 10^{-8}$   \\
         \hline
   Training steps per parameter ($T_{\text{train}}$)  &  3000         &  15000              \\
   \hline
    Transient steps ($T_{\text{trans}}$)  &          1000      &    1000            \\
        \hline
\end{tabular}
\end{center}
\end{table}

\section{Conclusion}

In this work, we demonstrated that a PARC scheme can effectively predict the dynamics and phase transitions of both a single logistic map and a two-dimensional coupled map lattice (CML). For the single logistic map, the reservoir was trained exclusively on bifurcation parameter values of $\mu$ before the attractor-merging crisis ($\mu_0 = 3.6786$). Despite having no data from the post-crisis regime, the PARC scheme accurately reconstructed the full bifurcation diagram and, more importantly, correctly identified the transition from an ordered phase ($M(\mu) \approx 0$) to a disordered phase ($M(\mu) > 0$) via the order parameter $M(\mu)$. This demonstrates that the reservoir learned the underlying parameter-dependent dynamics sufficiently well to extrapolate across a critical transition point.

Extending the approach to the high-dimensional CML, we exploited the spatial homogeneity of the lattice by training a single reservoir on the dynamics of just a few lattice sites. This strategy drastically reduces training data requirements and computational costs compared to conventional multi-reservoir or fully connected single-reservoir frameworks~\cite{Pathak2017, lei2026symmetry}. The trained reservoir successfully predicted the transition from in-phase synchronization ($\theta \approx 1$) to anti-phase clustered states ($\theta \approx 0$), again using only pre-transition training data. These results confirm that PARC, combined with symmetry exploitation, provides an efficient and accurate tool for forecasting phase transitions in spatiotemporal systems with minimal computational resources.

While the present study focused on logistic maps with fixed coupling strength, several directions remain for future work. First, extending the PARC framework to simultaneously handle variations in both the bifurcation parameter $\mu$ and the coupling strength $\varepsilon$ would enable prediction of full two-parameter phase diagrams~\cite{Kong2021}. Second, applying the symmetry-replication idea to heterogeneous or disordered lattices (e.g., quenched parameter noise) could test the robustness of the approach in more realistic settings~\cite{Sabe2024}. Third, integrating uncertainty quantification by using the distribution of predictions across multiple reservoir realizations would provide confidence intervals for critical transition forecasting. Finally, experimental validation on physical spatiotemporal systems, such as coupled electronic oscillators, would be a crucial next step toward practical deployment of PARC-based early warning systems for critical transitions.

\end{document}